\documentclass[12pt]{article}
\usepackage{graphicx}
\begin{document}
\centerline{How many different parties can join into one stable government ?}

\bigskip

Dietrich Stauffer

Institute for Theoretical Physics, Cologne University, 

D-50923 K\"oln, Euroland

\bigskip
Abstract: Monte Carlo simulations of the Sznajd model with bounded confidence
for varying dimensions show that the probability to reach a consensus in 
$d$-dimensional lattices depends only weakly on $d$  but strongly on the number
$Q$ of possible opinions: $Q=3$ usually leads to consensus, $Q=4$ does not.

\bigskip
In democracies where not just two parties dominate in elections, the government
often is formed by a coalition of several parties, since no single party won
more than 50 percent of the seats in parliament. According to (West) German 
experience of the last half-century, coalitions with up to three parties appear
often, those with four and more happen only rarely. Obviously, the more parties
a government has, the less stable it is in general. However, if parties need 5
percent of the vote to get into parliament (as is usually the case in Germany),
the total number of parties is limited anyhow, and the above rule of up to
three parties per coalition could simply come from the fact that typically
only five are represented. Thus we want to find a simple computer model to 
check if a consensus is indeed difficult to reach if more than three parties
try to form a government. 

The Sznajd model \cite{sznajd,jasss} (see \cite{aip} for a recent review) has
been shown to agree well with election statistics in Brazil and India 
\cite{bsk,gonzales} and thus is a natural choice for the present question.  We 
assume that there are $Q$ different opinions $q = 1,2, \dots Q$ (= parties) 
possible for each 
of $L^d$ sites (= politicians) on a hypercubic lattice in $d$ dimensions; if and
only if two neighbouring sites have the same opinion $q$, they convince their 
$4d-2$ neighbours to join party $q$. However, politicians are assumed to
switch only to 
politically neighbouring parties, i.e. from opinion $q \pm 1$ to opinion $q$: 
bounded confidence \cite{krause,deffuant}. The initial opinions are randomly 
distributed. Random sequential updating and free 
boundary conditions are used throughout; neighbours are always nearest 
neighbours. Opinions 1 and $Q$ are not taken as neighbours. The program 
sznajd31.f is available from stauffer@thp.uni-koeln.de.

\begin{figure}[hbt]
\begin{center}
\includegraphics[angle=-90,scale=0.43]{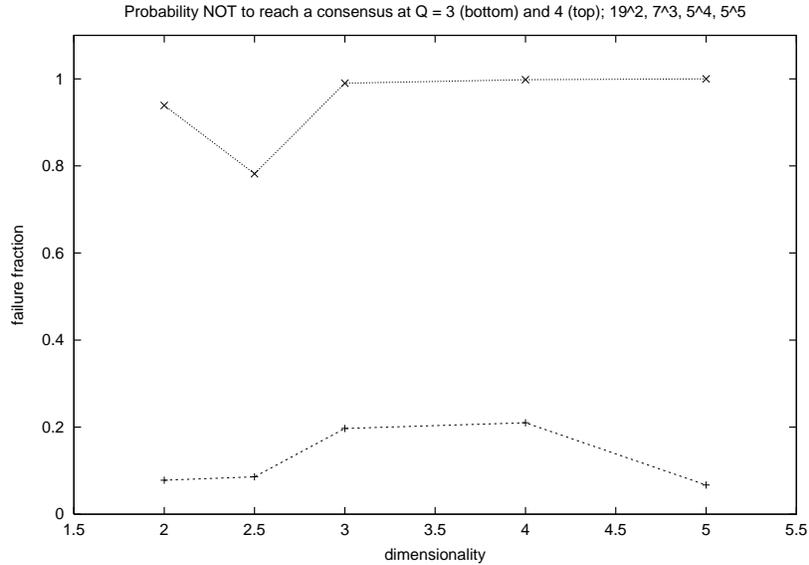}
\end{center}
\caption{Variation with dimensionality $d$ of the fraction of cases where no 
consensus is reached. The upper data correspond to $Q=4$ parties (which usually
fail), the lower to $Q=3$ parties which usually keep together. For larger 
lattices at $Q=3$ the failure probability goes to zero.
}
\end{figure}

The number of political leaders is much more limited than the number of voters,
and thus we work with rather small lattices, like $L = 19$, 7, 5 and 5 in 
two to five dimensions. Simulations are stopped if after 10,000 sweeps through 
the lattice no consensus is reached, if a fixed point without consensus is 
reached, or if
a consensus is reached with all $L^d$ sites having the same opinion $q$. Fig.1
shows that consensus is the rule for $Q = 3$ but rare for $Q = 4$, from 1000
separate simulations for each point. ($d = 2.5$ corresponds to the 
two-dimensional triangular
lattice with six neighbours for each site. With $Q=2$, always a consensus was 
found, which does not agree with the break-up of the West Germany federal 
government in 1982; India is a counterexample where many more parties are mostly
kept together since years in one government.)  This transition from consensus 
($Q=3$) to no consensus ($Q=4$) is similar to that in \cite{deffuant,redner}.

[Ref. \cite{jasss} claimed that for the triangular lattice the border 
between consensus and no consensus is shifted to $Q=5$; however, there a
consensus was also counted if the opinions separated into a fixed point with 
well separated opinions, like only $q = 1$ and 3 for $Q=3$: ``agree to
disagree''. Our figure now shows that indeed a true consensus for $Q=4$ is 
easier for the triangular lattice than for 
the square or simple cubic lattices, but still failures occur much more often
than consensus even on the triangular lattice.]

Thus not just the limited total 
number of parties in parliament is responsible for keeping limited the number
of parties in a government; it is also very difficult to reach a consensus 
among four parties.

\bigskip
This work started with a question of J. Liu from Harvard at \cite{aip} and
profitted from comments by J. Kert\'esz, D. Chowdhury and G. Weisbuch.

\end{document}